# The Fate of Frozen Carbonated Water at Europa-like Conditions

Swaroop Chandra , William T. P. Denman[2] , and Michael E. Brown
Division of Geological and Planetary Sciences, California Institute of Technology, Pasadena, CA 91125, USA; sc0296@caltech.edu

## Abstract

We present the results of experiments probing the retention of $CO_2$ in crystalline water ice, frozen sodium chloride (NaCl) brines, and flash-frozen carbonated water using diffuse reflectance infrared spectroscopy. Characteristic absorptions alluding to the formation of clathrate hydrates in crystalline ices and frozen brines are observed. NaCl in frozen brines does not qualitatively affect the formation of clathrate hydrates. Generation and stability of clathrates in crystalline ice transiently subjected to pressure–temperature ($P$–$T$) conditions in the stability region is observed, despite conditions being unviable at the onset of freezing. Retention of $CO_2$ in flash-frozen carbonated water is observed to be dependent on the temperature of the substrate during freezing. The state of $CO_2$ retained in the resulting ices differs from clathrate hydrates, as inferred from the respective infrared spectra. Both mechanisms of $CO_2$ retention are stable up to 140 K and under evacuated conditions. In the context of Europa, the $P$–$T$ states traversed by the samples plausibly represent the typical conditions around endogenous $CO_2$, if it is indeed transported from the subsurface ocean to the surface while being retained in ice/frozen brines and/or liquid emerging on the surface. However, the absorptions of $CO_2$ in laboratory infrared spectra do not match those detected on the leading side of Europa by the NIRSpec instrument on board JWST. Therefore, it is unlikely that the endogenous $CO_2$ observed at the surface of Europa is sourced directly from the ocean, unless additional processes affect the observed bands of $CO_2$ on Europa.

*Unified Astronomy Thesaurus concepts:* Ice spectroscopy (2250); Ice composition (2272); Surface ices (2117); Europa (2189); Infrared spectroscopy (2285)

## 1. Introduction

The presence of solid $CO_2$ on the surface of Europa was originally detected by the Near-Infrared Mapping Spectrometer (NIMS) on board the Galileo spacecraft via the presence of a weak band near 4.26 $\mu$m, corresponding to the $\nu_3$ asymmetric stretch of the molecule (T. B. McCord et al. 1998; G. B. Hansen & T. B. McCord 2008). The stability of $CO_2$ at the 90–130 K surface temperatures of Europa (J. R. Spencer et al. 1999) is surprising, given that crystalline $CO_2$ sublimes at temperatures exceeding 70 K under ultrahigh vacuum conditions prevailing on the surface of Europa (C. E. I. Bryson et al. 1974). The continued presence of $CO_2$ under these conditions thus indicates that the $CO_2$ is trapped by some less volatile material (R. W. Carlson et al. 2009). Recent higher spectral resolution and higher signal-to-noise ratio observations of the leading hemisphere of Europa by the NIRSpec instrument on board JWST resolved the $\nu_3$ band into a doublet profile with band centers at 4.249 and 4.269 $\mu$m. Additionally, the $\nu_1 + \nu_3$ combination band at 2.695 $\mu$m was detected (S. K. Trumbo & M. E. Brown 2023; G. L. Villanueva et al. 2023). Mapping the concentration of $CO_2$ across the surface of Europa also revealed its increased abundance in chaos terrains, with the maximum located at Tara Regio. Some of the proposed mechanisms of formation of the chaos terrains involve exchange of material between the inferred salt-rich subsurface ocean (J. D. Anderson et al. 1998; M. G. Kivelson et al. 2000) and the ice shell. The increased abundance of $CO_2$ observed at chaos terrains could thus indicate an internal carbon source. An estimate of the lifetime of $CO_2$ at the equatorial regions of Europa by a Monte Carlo model incorporating sublimation, ballistic transport, and loss of $CO_2$ by photoionization indicates the requirement of an ongoing supply of $CO_2$ at the surface (S. Kadoya et al. 2025). Therefore, $CO_2$ could be the product of radiolytic and/or chemical processing of carbon-based material sourced from the ocean, or the $CO_2$ itself could be sourced directly from the ocean.

Release of $CO_2$ and other volatiles such as $O_2$, $SO_2$, $CH_4$, $N_2$, and $H_2S$ into the ocean has been hypothesized for icy ocean worlds (T. M. Becker et al. 2024). Some of the processes include degassing of $CO_2$ from carbonate/bicarbonate-based minerals in the ocean (M. Melwani Daswani et al. 2021), breakdown of dry ice shelves on the seafloor (N. C. Shibley & G. Laughlin 2021)—plausible at the postulated pressures of ~1500–1700 bars (G. M. Marion et al. 2003; S. Wakita et al. 2024), its presence in hydrated forms such as carbonic acid and/or clathrates, and even infusion of primordial $CO_2$ during later parts of differentiation (O. Mousis et al. 2023; J. M. Weber et al. 2023). Leaching of carbon from the silicate seafloor through serpentinization and hydrothermal activity are also processes that release $CO_2$ alongside $CH_4$ into the ocean. These reactions could mirror those seen in Earth's hydrothermal vents, where water–rock interactions contribute to carbon cycling (S. D. Vance et al. 2016). Formation of clathrate hydrates of the volatiles following their release into the ocean is plausible, given their stability under the lithostatic pressure and temperature conditions in the interior of Europa (O. Prieto-Ballesteros et al. 2005; A. D. Fortes & M. Choukroun 2010; F. Sohl et al. 2010; M. Choukroun et al. 2013; A. Bouquet et al. 2019). Depending on the density of the brine ocean, layers of clathrates are expected to sink to the seafloor or float, becoming pivotal factors affecting continuation/termination of mantle–ocean interactions. The $CO_2$ thus forming a constituent of the ocean could permeate and be retained within ice at the ocean–ice interface beneath the ice shell of Europa.

---

[2] Present Address: School of Chemistry and Biochemistry, Georgia Institute of Technology, Atlanta, GA 30332, USA.

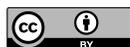

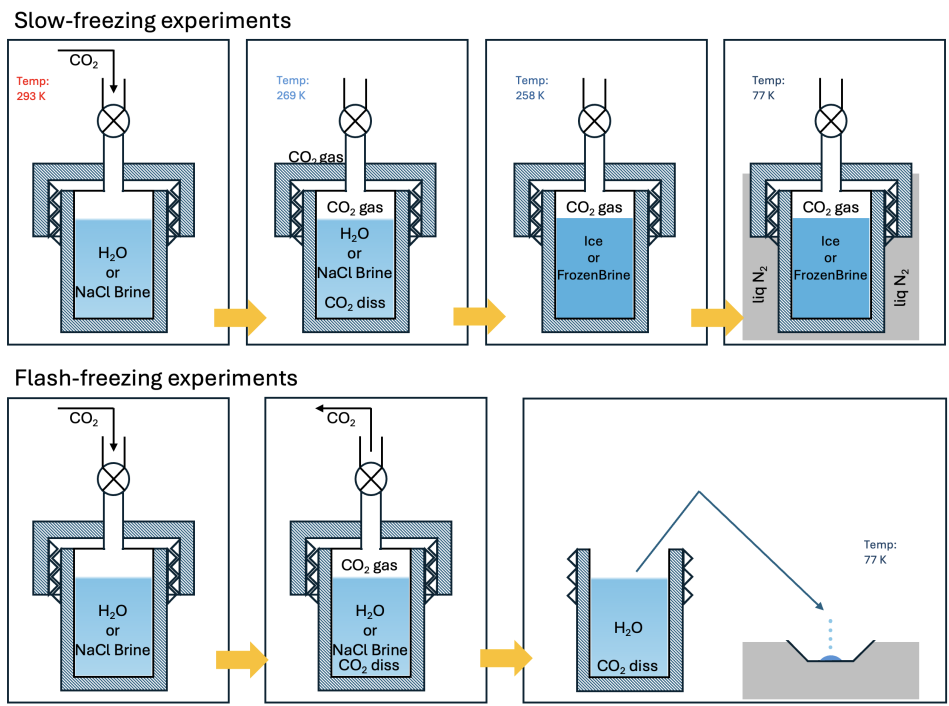

**Figure 1.** Schematic of the procedure in two sets of experiments performed with carbonated water.

The emergence of $CO_2$ on the surface while retained in ice, as the latter migrates to the surface, forms a potential sourcing mechanism of endogenic $CO_2$ on Europa. Rapid freezing of the liquid phase of the ocean at the surface or as inclusions at shallower depths from the surface are also possible during proposed cryovolcanic activity (G. D. Crawford & D. J. Stevenson 1988; J. S. Kargel 1991; L. Wilson et al. 1997; L. C. Quick et al. 2017). The existence of endogenous $CO_2$ remains the sole unambiguous evidence for the availability of carbon, a biologically essential element, on Europa. Identifying the chemical forms of carbon in the ocean enables constraints on its redox state and pH, which in turn could inform of the origins and formation of Europa, as well as have implications for its habitability. If it is indeed $CO_2$ from the ocean which is transported to the surface, the ocean would have to be slightly acidic (M. A. Pasek & R. Greenberg 2012), in support of slow evolution of the interior with preservation of volatiles (M. Melwani Daswani et al. 2021). The presence of an acidified ocean would also point to an oxidized ocean with potential excess of oxidants. Based on the models of M. A. Pasek & R. Greenberg (2012) and retrospection, the presence of dissolved $CO_2$ could also be evidence in favor of the delivery of material into the ocean from the ice shell. It also enables better constraints on the present-day pH of the ocean, which in turn allows the deduction of when its potential oxygenation could have commenced. The pH of the ocean further constrains the composition of the seafloor: An acidic ocean is indicative of ineffective neutralization through rock–water interactions, providing insights into seafloor composition, such as silicate-bearing basalts and/or nonreactive sulfates. Conversely, a basic ocean, which would be indicated by the speciation of oxidized inorganic carbon as $HCO_3^-/CO_3^{2-}$ rather than $CO_2$, would suggest the opposite scenario, in which rock–water interactions involving the seafloor release cations such as $Na^+$, $K^+$, $Ca^{2+}$, and $Mg^{2+}$ into the ocean. In short, evidence for the presence or absence of dissolved $CO_2$ in the ocean can constrain the types of constituent electrolytes, the surfaces available for heterogeneous catalysis driving metabolic reactions, and the aerobic or anaerobic nature of that potential biology that could be sustained in the ocean, thus furthering our understanding of the ocean's habitability.

In this work, we perform laboratory experiments intended to simulate aspects of the incorporation of dissolved $CO_2$ into solid ice on Europa and the emergence of the materials on the surface. We consider two separate processes. The first process represents the retention of $CO_2$ in ice at the ocean–ice interface beneath the ice shell of Europa, following which it is transported to the surface during the migration of ice. The second process represents the flash freezing of liquid from the ocean on direct exposure at the surface of Europa. We examine the retention and state of $CO_2$ via diffuse reflectance infrared spectroscopy. Finally, we discuss implications for the presence of $CO_2$ on Europa.

## 2. Experimental Methods

### 2.1. Sample Preparation

Our experiments (Figure 1) attempt a qualitative replication of the possible journey of $CO_2$ from the ocean to the surface. In one set of experiments, we simulate the dissolution of $CO_2$ into ocean water, the slow solidification of ocean water at temperatures just below the freezing point, the cooling of the ice to the surface temperature of Europa as it rises to the surface, and, finally, the exposure at the surface. For these experiments, we use both pure water (from Milli-Q ultrapure water) and NaCl brines as our simulated ocean water. In our second set of experiments, we replicate the much simpler process of ocean water with dissolved $CO_2$ being directly exposed to the surface of Europa. The full procedure for both sets of experiments is described below. The varying pressure–temperature ($P$–$T$) conditions within the vessel for five separate experiments are traced on the overlaid phase diagrams of $CO_2$ and $H_2O$ in Figure 2.

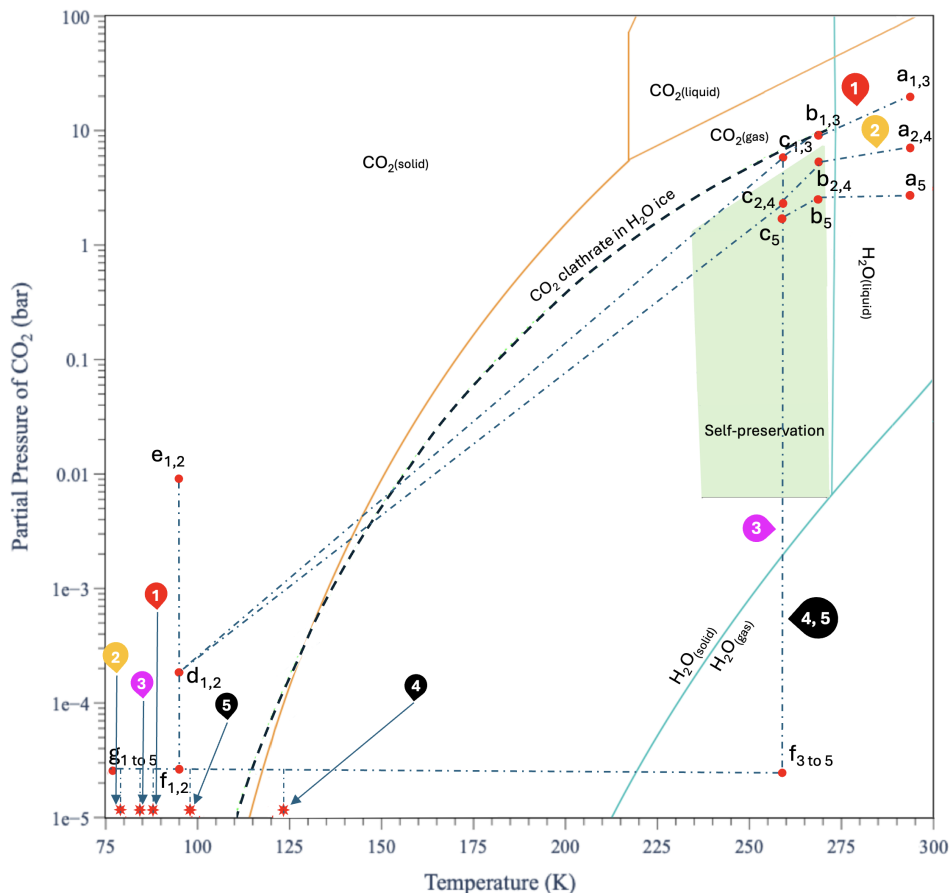

**Figure 2.** Pressure–temperature ($P$–$T$) conditions of our experiments compared to phase diagram of the $CO_2$–$H_2O$ system. Red dots correspond to actual $P$–$T$ conditions at various stages of ice preparation. Dashed lines are only indicative and do not correspond to measured variations in pressure. Paths labeled in red, yellow, and magenta correspond to ice retaining $CO_2$. Paths labeled in black correspond to ice devoid of $CO_2$. Lowercase letters show the stages of the experiment described in the text, with subscripts showing the applicable experiment number. The stars near the bottom left denote the $P$–$T$ conditions during acquisition of spectra.

(a) A volume of 20 ml of the liquid phase (either pure water or brine) is carbonated by pressurization with $CO_2$ in a high-pressure vessel (25 ml/SS T316/Series 4740, Parr Instrument Company). The extent of carbonation is estimated by means of pH strips placed in the liquid phase.

(b) In the first set of experiments (Figure 2), the pressurized liquid phase is cooled to 269 K in a refrigerator for 1 hr, maximizing the solubility of $CO_2$. This step approximates the coldest liquid at the ocean–ice interface of Europa.

(c) For the first set of experiments, the pressure vessel is moved to an adjacent freezer maintained at 258 K for 1 hr, where the ice/frozen brine forms. The resulting solid phase approximates the warmest ice/frozen brine at the ocean–ice interface. For the second set of experiments (Figure 3), the $CO_2$ pressure is released and the liquid is dispensed drop-wise with a pipette onto a stainless steel mortar cooled with liquid $N_2$, approximating the effects of flash freezing. For these experiments the pressure of $CO_2$ is always 2.5 bars. After flash freezing we immediately proceed to step (g).

(d) The vessel is then cooled by immersion in liquid $N_2$ for about 20 minutes. This step represents the cooling of the solid phase as it migrates to the surface.

(e) The vessel is moved to a positive-pressure glove box purged with dry $N_2$, maintaining relative humidity levels below 1%.

(f) Any residual pressure of $CO_2$ is released, after which the screw-cap head seal of the vessel is disengaged.

(g) For the first set of experiments, a sample of the ice core, at about 10 mm depth from the top, is ground in a liquid $N_2$-cooled stainless steel mortar and loosely packed (with no application of any compressive force) into a precooled sample holder. The sampling is done at 10 mm depth to avoid inclusion of any condensed $CO_2$ at the top of the core formed during the liquid nitrogen cool-down. For the second set of experiments, the frozen material is directly ground on the cryogenic mortar and transferred to the sample holder. The grain size is unconstrained because grain sizes do not affect the band center of the absorption lines of concern in the study. This also enables rapid transfer of the grains to the precooled sample holder for spectral acquisition. The sample holder is part of a Praying Mantis™, low-temperature reaction chamber (CHC-CHA-4, from Harrick Scientific, henceforth "the chamber") equipped for spectroscopy under evacuated conditions, enabling spectroscopy under *P*–*T* conditions expected on the surface of Europa. More details on the equipment are presented in the following subsection. Grinding represents the geological processing (R. N. Clark et al. 1983; M. L. Nelson et al. 1986) that the ice could be subjected to during migration and/or at the surface of Europa.

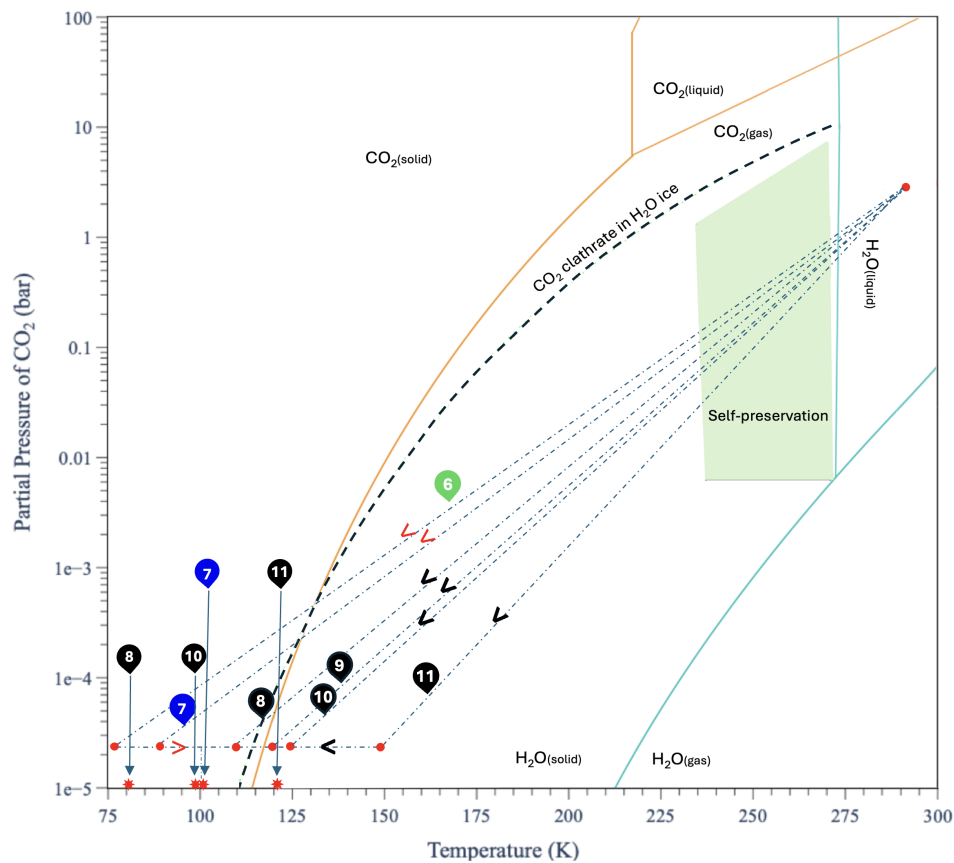

**Figure 3.** Phase diagram of the $CO_2$–$H_2O$ system. Red dots correspond to actual *P*–*T* conditions at various stages of ice preparation. Dashed lines are only indicative and do not correspond to measured variations in pressure. Paths labeled in blue and green correspond to ice retaining $CO_2$. Paths labeled in black correspond to ice devoid of $CO_2$. The stars near the bottom left denote the *P*–*T* conditions during acquisition of spectra.

The temperature at the top of the ice core in steps (d) to (f) is not determined for each experiment, as no temperature probe is placed within the pressure vessel. Instead, the temperature is estimated from a separate nonpressurized control run where we placed a PT100 class A RTD probe in contact with water (20 ml) within the vessel, such that its tip was immersed at a depth of ∼3 mm. The probe was suspended such that it had mechanical contact only with water and not with the sides of the vessel. The vessel was then cooled to 258 K and then immersed in liquid $N_2$, while the drop in temperature was recorded at intervals of 30 s until thermal equilibrium was reached at 126 K. This is treated as the temperature at the top of the ice core, which is in contact with the residual $CO_2$ gas in the head space within the vessel. W. F. Giauque & C. J. Egan (1937) report an empirical equation relating the vapor pressure of $CO_2$ with the temperature obtained from a fit of their vapor-pressure/temperature measurements. This equation is used to calculate the pressure of $CO_2$ in the head space within the vessel in steps (d) and (e). A separate RTD probe was used to measure the temperature at the sampling depth of 10 mm, where an equilibrium temperature of 90 K was found. The warming at 10 mm depth is negligible during sampling, as calculated using the simple one-dimensional diffusion of heat equation for ice. The surface of the ice core rises to about 150 K during the time that it takes to release the pressure from the vessel. The upper limit to the head-space pressure during this period is calculated from the vapor pressure at 150 K (step "e" on Figure 2). At this stage any residual pressure of $CO_2$ from the head space within the vessel is released, denoted by the vertical drop in pressure from (e) to (f). The final pressures shown (in steps "f" and "g") are estimated upper limits to the $CO_2$ partial pressure in the glove box and the chamber, respectively. We calculate these by assuming a linear correspondence in the reduction of moisture and $CO_2$ within the glove box, and deduce the partial pressure of $CO_2$ from the measured humidity within the glove box (reduced and maintained at <1% from an ambient humidity of 35% by purging $N_2$). Finally, the sample is ground within the liquid $N_2$-cooled mortar, which is represented by point "g" on the figure.

In our first two trials, water was pressurized to 20 and 7 bars of $CO_2$, corresponding to the initial conditions of paths 1 and 2, respectively, in Figure 2. A pressure of 20 bars is the upper limit achievable within the vessel, and should be sufficient for enclathration of $CO_2$ upon freezing (A. Y. Manakov et al. 2009). A pressure of 7 bars ensures that the initial freezing does not cause enclathration. The resulting solutions should have $CO_2$:$H_2O$ ratios of 25:1000 (the ratios correspond to mole fractions of $CO_2$ per 1000 moles of water) at 20 bars/259 K and 9:1000 at 7 bars/259 K, calculated using the empirical solubility model reported by J. J. Carroll et al. (1991) for water at 273 K. Following pressurization, a drop in pH from 6.7 to <2.5 at 7 and 20 bars was observed using the pH strips.

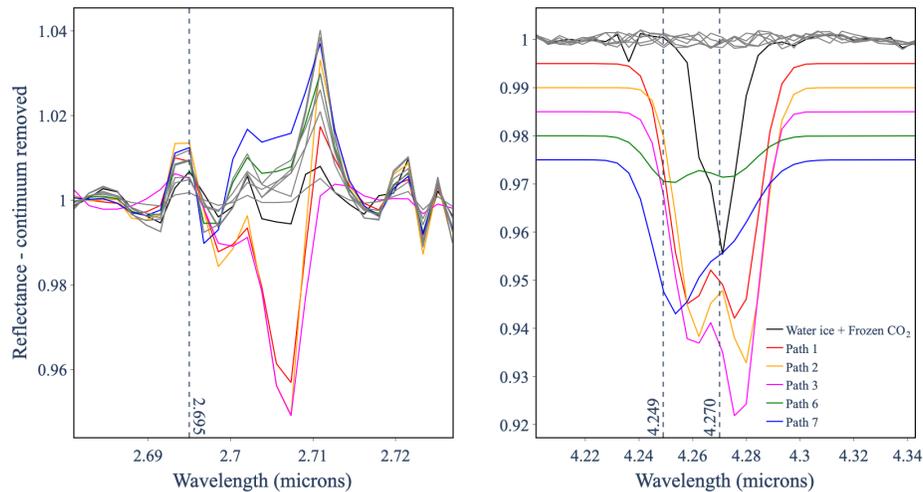

**Figure 4.** Infrared spectra of ices after pressurization with $CO_2$. Also shown, in black, is a spectrum of pure frozen $CO_2$ mixed with water ice, at a temperature of 125 K. Traces in gray correspond to paths 4, 5, and 8–11. No visible retention of $CO_2$ is observed in these instances. The dashed lines mark the absorptions observed on the leading side of Europa. The absorption at 4.249 μm is deeper than that at 4.270 μm, in the disk-integrated near-IR spectrum of Europa's leading side.

All NaCl brines (5%, 10%, and 23.3% w/v; 23.3% w/v corresponds to the eutectic of the NaCl–$H_2O$ two-component system; P. V. Cohen-Adad & J. W. Lorimer 1991) were pressurized with 7 bars of $CO_2$, corresponding to path 2 in Figure 2. Brines at 5% and 10% w/v required about 2 hr at 258 K to freeze over completely, compared to 1 hr for water. However, the drop in pH was similar to the corresponding trials with water. The brine at the eutectic concentration of 23.3% w/v does not freeze at 258 K (eutectic point 252 K).

### 2.2. Infrared Spectroscopy

The chamber is equipped with a three-port dome, two of which are fixed with ZnSe windows and the third with a quartz window. The volume within the dome, including the sample holder, is evacuated to a pressure of ∼0.03 bar (20 Torr). The sample holder is precooled with liquid $N_2$, filled into the dewar of the chamber. At this pressure, the sample holder reaches an equilibrium temperature of 100 K. The Praying Mantis is equipped with two 90° off-axis ellipsoidal mirrors, one of which focuses the incident infrared beam onto the sample, while the other collects the diffuse reflection from the sample across a large solid angle (accounting for ∼20% of reflected light).

Each spectrum is an average of 256 scans collected by a Nicolet iS50 FT-IR spectrometer equipped with a liquid $N_2$-cooled MCT-A (mercury cadmium telluride, semiconductor) detector, operating across the range 8000 cm$^{-1}$ to 1000 cm$^{-1}$ (1.25 μm to 10.00 μm) at a resolution of $\Delta\tilde{\nu} = 0.241$ cm$^{-1}$ ($R = \frac{10,000}{\lambda \cdot \Delta\tilde{\nu}} - 1$; $R$ decreases from ∼33,200 at 1.25 μm to ∼6900 at 10.00 μm as a function of $\lambda$). The entire beam path within the spectrometer and the Praying Mantis is purged with $N_2$ at all times to minimize telluric absorption. The chamber is wrapped in cellophane to effect a partial seal on the volume within, minimizing the amount of ambient $CO_2$. All spectra recorded are ratioed with that of an Infragold (Labsphere) standard, recorded under similar conditions. Five trials of every P–T path of ice preparation are performed. Likewise, five trials of preparing frozen brines at every concentration are undertaken. The five spectra for a given path/concentration are averaged to generate the spectra displayed and used for analysis.

Measuring the temperature of macroscopic samples cooled from below and radiatively heated from above using an embedded thermocouple is unreliable. We thus determine the true temperature of ice directly from the infrared spectrum. G. Filacchione et al. (2016) show that the wavelength of the ∼3.5 μm peak in reflectance of water ice between the 2.6 and 3.6 μm O–H stretch and the 3.6–5.0 μm $\nu_2 + \nu_L$ combination band shifts with the temperature of the water ice. We fit a third-order polynomial to the reflectance across 3.4–3.8 μm to determine the position of the maximum. It should be noted that the correlation reported by G. Filacchione et al. (2016) is limited to temperatures ranging across 88–172 K. We presume that this correlation continues to hold even for colder ices, at least down to 80 K. The presence of NaCl in the brines also leads to a shift in the position of this peak, so no precise temperatures are reported for these results, but they should be within the same range.

To examine the absorptions of $CO_2$ in the 2.7 and 4.2 μm regions, the reflectance across 2.68–2.73 and 4.2–4.35 μm was ratioed to a third-order polynomial fit, representing the continuum between the respective bounds. Within the range 4.2–4.35, data points between 4.23 and 4.29 were excluded from the fit. Likewise, in the range 2.68–2.73, points between 2.7 and 2.714 were excluded.

### 3. Results and Discussion

#### 3.1. Retention of $CO_2$ in Ice/Frozen Brines Formed by Slow Freezing

The spectra of the samples following paths 1 and 2 using pure water for the liquid phase are shown in Figure 4. In these cases a $CO_2$ doublet line is observed, with centers at 4.258 and 4.278 μm, accompanied by a relatively weak absorption at 2.706 μm. These absorptions match those of clathrate hydrates of $CO_2$ in ice, as reported by A. Oancea et al. (2012). As the sample is not always within the clathrate formation region or within the known self-preservation region (Figure 2), clathrate hydrates were not anticipated to be present in the final sample during these experiments.

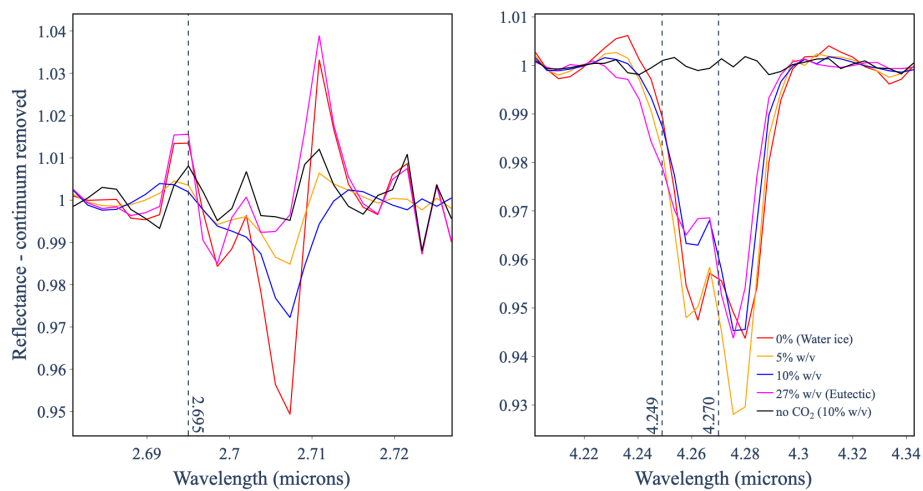

**Figure 5.** Infrared spectra of frozen brines produced following pressurization of brines with 7 bars of $CO_2$. The dashed lines mark the absorptions observed on the leading side of Europa. The absorption at 4.249 μm is deeper than that at 4.270 μm in the disk-integrated near-IR spectrum of Europa's leading side.

In the highest-pressure experiments (path 1), clathrate hydrates would have formed and remained stable within the ice formed at 258 K. They were apparently preserved during the cooling with liquid $N_2$. The P–T conditions during grinding in the cold mortar (at 77 K, with relative humidity <1%) within the glove box fall within the stability region. The P–T state in the chamber, at temperatures ranging from 80 to 100 K and evacuated to 0.03 bar (total pressure), is also within the stability region. The sample was maintained at these conditions for 10 hr, during which the doublet absorptions and the feature at 2.706 μm persisted with no loss of intensity.

The presence of clathrate hydrate in the lower-pressure case (path 2) is surprising, as at this lower pressure the clathrate hydrate should not form during the initial freezing. It is unclear where in this process the clathrate hydrate would first appear. Additional experiments (paths 3–5; described below) were performed to attempt to find the answer to this question.

All experiments with NaCl brines were performed using the same P–T conditions as path 2. The spectra obtained from these experiments (Figure 5) show $CO_2$ absorptions at the same wavelengths as the pure-water experiments, possibly due to the near-complete rejection of NaCl from the lattice of water ice. However, the intensity of absorption at 2.706 μm is visibly diminished in cases with 5% and 10% concentrations of NaCl. The 2.706 μm disappears completely in the case of the brine at eutectic concentration, but this case is unique, as the liquid phase would not freeze at 258 K. At this point, it appears that, regardless of the P–T conditions at which the liquid phase first freezes, the brief passage of the resulting ice through the stability region of clathrate hydrates is enough for their formation.

In an attempt to understand where clathrate hydrate formation occurred during the lower-pressure path 2 experiments, ices were prepared via three new pathways (paths 3, 4, and 5; see Figure 2). Pathways 3 and 4 are analogous to paths 1 and 2, respectively, in that the initial pressures of $CO_2$ over water are identical (20 and 7 bars, respectively). However, the pressure of $CO_2$ is released immediately after the formation of ices at 258 K rather than after the cool-down under liquid $N_2$. The warm-up during transfer to the glove box here is negligibly small compared to that in paths 1 and 2. A sample of ice is retrieved and ground in liquid $N_2$-cooled mortar and then packed into the chamber. An identical experiment (path 5) was also performed with an even lower $CO_2$ pressure of 2.5 bar. As seen in Figure 4, ice from the high-pressure path 3 experiment—in which the pressure is high enough that the clathrate can form at the time of the initial freezing—retains $CO_2$, whereas no absorption due to $CO_2$ is detected for ices from paths 4 and 5.

We conclude from these experiments that the formation of clathrate hydrates in the case of path 2 occurs at a stage after the formation of ice at 258 K. The release of $CO_2$ pressure at 258 K appears to enable its complete expulsion from the solid phase, while maintaining pressure during cooling seems to facilitate some retention. It is known that the instant subcooling of a mixture induces nucleation of hydrates. Subcooling indicates the rapid decrease of temperature from the equilibrium temperature. The rate of induction and propagation of the formation of hydrates is proportional to the degree of subcooling, with the lowest margin required being 3.6 K ($T_{eq}$–T; A. K. Sum et al. 2009). The occurrence of nucleation at gas–liquid and gas–liquid/container wall interfaces is explained by C. P. Ribeiro & P. L. C. Lage (2008). The gas–liquid interface is found to be the primary site of nucleation for gases such as hydrocarbons, which are less soluble in water. For $CO_2$, however, the gas–liquid/container wall interface is found to be the primary site of nucleation given the mutual solubility of $CO_2$ in water as well as water in $CO_2$. The walls of the pressure vessel rapidly cool to 77 K from 258 K on immersion into liquid nitrogen. As demonstrated by S. Takeya et al. (2000) for $CO_2$, this cooling under pressurized $CO_2$ facilitates the formation of a layer of liquid $CO_2$ that is simultaneously in contact with the walls of the pressure vessel and a "thin" film of liquid water above the water ice, initiating the nucleation of hydrates. However, a release of pressure before immersion would not facilitate this rapid nucleation. This could be an explanation for the surprising presence of clathrate hydrates in path 2. In short, clathrate hydrates produced at 258 K are preserved through subsequent steps in path 1, whereas in path 2 their formation occurs sometime during cooling with liquid $N_2$ and/or as the pressure of $CO_2$ is released.

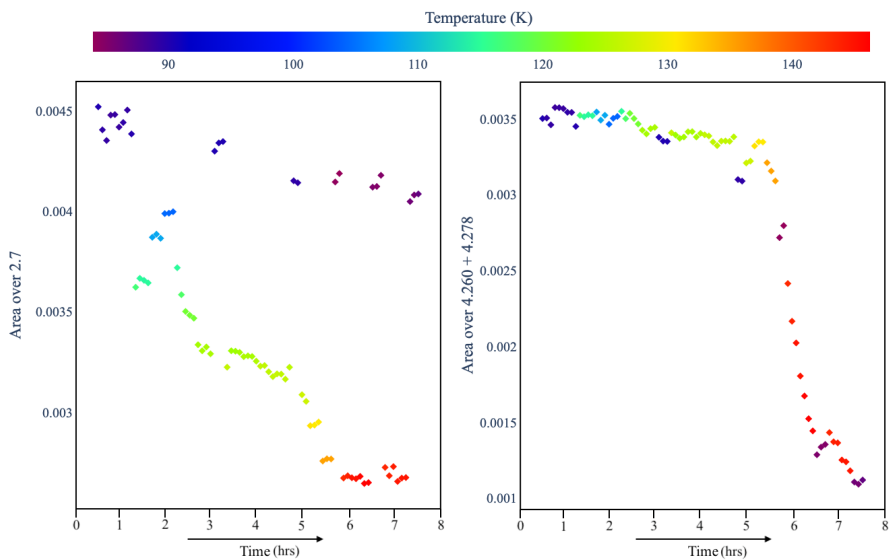

**Figure 6.** Change in intensity of absorptions at 2.706 $\mu$m (left) and 4.258/4.278 $\mu$m (right) with temperature. The abscissa is time. The reversible change in intensity of 2.706 $\mu$m is evident from the consistently greater magnitudes of area corresponding to the blue/violet points in the left panel. The stability of the doublet absorption up to 140 K and the irreversible drop in intensity at greater temperatures is seen in the right panel.

To check whether the stability of clathrates is transient, given the short time the sample spends in these regions before reverting to the stability region during grinding, the ices were progressively warmed up to 150 K in the chamber. Beyond 150 K, the ice/frozen brine began to sublimate. The intensity of the doublet absorption remained unchanged up to 140 K, beyond which the intensity began to decrease. This drop was irreversible after cooling (Figure 6). During thaw–cool cycles, the pressure in the reaction chamber remained at 0.03 bar. The absorption at 2.706 $\mu$m diminished in intensity as warming progressed but reversed upon cooling. The absorption regained intensity, by as much as 89% of its original value, upon cooling to 100 K, even after warming to 150 K. In short, the doublet absorption proved stable up to 140 K, while the intensity of absorption at 2.706 $\mu$m varied with temperature and remained stable at least up to 150 K. Thus, the stability of hydrates under these P–T conditions, outside the region of stability up to 140 K, is established. Clearly, the self-preservation region extends beyond that marked in Figure 2 based on the observations of A. Oancea et al. (2012), at temperatures between 115 and 150 K.

The possibility of an extended region of self-preservation seems plausible, based on the hypothesis of V. R. Belosludov et al. (2018). Their models predicted an increased relative thermal expansion of structure I (sI) clathrate hydrate sites occupied by $CO_2$ compared to empty sites. The modeled magnitudes of relative thermal expansion of sI sites was in good agreement with those observed experimentally. When the temperature of a clathrate hydrate is increased beyond its region of stability but remains below the freezing point of water ice, models predict the generation of excess pressure within the clathrate hydrate domains dispersed in water ice. This pressure comes from the differential thermal expansion between clathrate hydrates and water ice, and is posited as the mechanism for self-preservation. A corollary is that the partial pressure of $CO_2$ above the ice does not affect the stability of clathrate hydrates, in turn allowing the extension of these results, obtained at 0.03 bar in the reaction chamber, to the surface of Europa, where the pressures are far lower. Therefore, if clathrate hydrates indeed formed on Europa, they could be preserved at the surface as well.

### 3.2. Retention of $CO_2$ in Ice Formed by Flash Freezing

$CO_2$ could be transported to the surface of Europa directly in liquid brines. Such brines would then be flash frozen when exposed to the surface. Additional experiments were performed to simulate these conditions. Water pressurized with 2.5 bar of $CO_2$ was flash frozen at temperatures from 77 to 150 K (paths 6–11; see Figure 3). Temperatures higher than 77 K were maintained using a slurry of liquid $N_2$ and dry ice. The quantity of liquid $N_2$ added determines the temperature at thermal equilibrium of the slurry controlled by the processes of cooling of dry ice with liquid $N_2$ and the heating/vaporization of liquid $N_2$ by dry ice. Ices obtained via paths 6 and 7 (corresponding to flash freezing of carbonated water at 77 and 90 K, respectively) retain $CO_2$, as indicated by doublet absorption with band centers at 4.252 and 4.271 $\mu$m. These absorptions are shifted to shorter wavelengths relative to those obtained for crystalline ices cooled by immersion in liquid $N_2$ under $CO_2$ pressure (paths 1 and 2). The weak absorption feature at 2.706 $\mu$m is absent in the case of ices formed by flash freezing. The absorption at 4.252 $\mu$m is more intense than that at 4.271 $\mu$m, which is the opposite of what is seen in the doublet of crystalline ice cooled while under $CO_2$ pressure. Ice from paths 8–11 (corresponding to mortar temperatures of 110, 120, 125, and 150 K, respectively) shows no measurable absorption due to retained $CO_2$, regardless of the final temperature. The doublet features from paths 6 and 7 remain stable for 10 hr in the chamber at 100 K and continue to be stable up to 140 K, as in the case of crystalline ice cooled under $CO_2$ pressure. Separate panels showing the intensity response of absorptions to temperature changes have not been included since those in Figure 6 display the exact same phenomenon. The temperature–intensity loci would only be offset from that in Figure 6 due to variations in the intensity of $CO_2$ absorption across experiments.

While the slow-freezing experiments produced crystalline water and subsequent clathrate hydrates, water ice formed by flash freezing would have a significant fraction of domains of hyperquenched glassy water (HGW; I. Kohl et al. 2000; R. M. E. Mastrapa et al. 2013). In experiments involving freezing of aerosols, a substrate temperature of 130 K or lower is reported to produce almost 100% HGW (I. Kohl et al. 2000). In our experiments, involving the freezing of small droplets (∼1 ml each), some fraction of crystalline ice is likely still present, albeit at much lower levels compared to those produced in the slow-freezing experiments. The fraction of crystalline ice likely decreases with lower temperature. We suspect that the $CO_2$ seen in the lowest-temperature flash-freezing experiments (paths 6 and 7) is trapped in HGW, whereas at higher temperature the ice is predominantly crystalline, which does not retain $CO_2$ upon flash freezing. In this interpretation, the difference in the $CO_2$ spectrum between the slow-freezing and flash-freezing experiments reflects the difference between $CO_2$ trapped as a clathrate and $CO_2$ trapped via a different mechanism in HGW. Note that amorphous solid water (ASW), which is likely not produced in these experiments, is also capable of trapping $CO_2$, but this trapped $CO_2$ has a single absorption at 4.275 $\mu$m at 95 K (M. B. Gálvez et al. 2008).

### 3.3. Concerning Europa

At the ∼36–460 bar pressures expected at the bottom of Europa's ice shell (N. S. Wolfenbarger et al. 2022; S. M. Levin et al. 2026), any $CO_2$ directly incorporated into the ice shell of Europa is likely to be stable as a clathrate hydrate (A. Y. Manakov et al. 2009). In addition to the pressures expected at the ocean–ice interface, pH and the presence of salts would influence the availability of $CO_2$ for the formation of clathrate hydrates. The dissolution of $CO_2$ in water is represented by the reaction $H_2O + CO_2 \rightarrow H_2CO_3$, with the generation of carbonic acid lowering the pH. This is seen in the experiments detailed above, with the drop in pH being proportional to the pressure of $CO_2$ applied. At 20 bars the pH is noted to be <2.5. Therefore, inhibiting the formation of hydrates by acidification so as to lower the dissolution of $CO_2$ would require acids stronger than $H_2CO_3$, as demonstrated by R. B. Lamorena & W. Lee (2009), where the presence of HCl delayed the formation of hydrates ∼50-fold. While the strong electrolyte HCl causes an effective pH of 2.19 due to ionic strength effects, the pH itself inhibits the formation of $H_2CO_3$, the state of dissolved $CO_2$ in water, by promoting the reaction $(H_2O + CO_2 \rightarrow H_2CO_3)$ leftward. The lowered availability of dissolved $CO_2$ in turn inhibits clathrate formation. Conversely, a pH >7 was seen to convert $CO_2$ to $CO_3^{2-}/HCO_3^-$ or cause rejection from the solution as carbonates, both lowering the availability of dissolved $CO_2$ for the formation of hydrates. However, the pH across the ocean and ice column of Europa is estimated to be between 4.5 and 5.5 by the models of M. Melwani Daswani et al. (2021). Therefore, inhibition of clathrate hydrate formation due to the lowered availability of $CO_2$ from pH effects seems unlikely at the ocean–ice interface of Europa.

The presence of ions such as $Na^+$, $Mg^{2+}$, $Ca^{2+}$, and $K^+$ is known to inhibit hydrogen bonding between water molecules due to the stronger ion–water interactions. This interferes with the formation of the characteristic cagelike structures of clathrate hydrates. This effect manifests as a depression in the formation and dissociation temperatures of the hydrates. However, the studies demonstrating these effects (E. D. Sloan & C. A. Koh 2007; K. M. Sabil 2009) involve significantly greater (10×) concentration of electrolytes compared to those expected at the ocean–ice interface of Europa (N. Schilling et al. 2007; M. Melwani Daswani et al. 2021). Therefore, the continued influence of electrolytes on hydrate formation at relevant concentrations is yet to be explored. Consequently, the incorporation of $CO_2$ as clathrate hydrates under the expected pressures at the ocean–ice interface is plausible. As experiments here have shown, such ices directly transported to the surface should contain the spectral signatures of clathrate hydrate. However, the double $CO_2$ absorption seen with JWST/NIRSpec does not occur at the same wavelengths as the doublet expected from clathrate hydrates. We thus conclude that this straightforward mechanism is not responsible for the $CO_2$ seen on Europa.

A second plausible pathway for $CO_2$ transport from the ocean is the movement of the liquid phase through the ice shell via cryovolcanism (L. Wilson et al. 1997; L. C. Quick et al. 2017). Subsequent flash freezing of the liquid phase occurs upon exposure at the surface. Generating ices through flash freezing of carbonated water simulates this scenario, while also enabling the explanation of the appearance of clathrate hydrates in frozen brines formed from NaCl solutions at eutectic concentrations. The doublet absorptions of ices produced by flash freezing differ markedly from those of clathrate hydrates in terms of their relative intensity and band centers. The absence of absorption at 2.706 $\mu$m adds to the conviction that the retained $CO_2$ in these ices does not correspond to clathrate hydrates. The stability of retention up to 140 K, in ices obtained by flash freezing at low enough temperatures, is reminiscent of the hypothesis of M. P. Bernstein et al. (2005). Their experiments involved codeposition of $CO_2$ and water vapor on a cold substrate. The stability of solid $CO_2$ within water at temperatures exceeding 90 K was attributed to highly localized domains of $CO_2$, preventing sublimation.

In summary, stable retention of $CO_2$ in water ice formed via both plausible pathways generates doublet absorptions; however, neither matches those observed on Europa. The retention mechanisms also remain unaffected up to 140 K under evacuated conditions in the chamber, suggesting they would also be stable at the surface of Europa. Nevertheless, these mechanisms alone appear incapable of explaining the presence of $CO_2$ on Europa. Modifications to the simple pathways explored here, such as the effects of radiation, alternative ocean chemistry, or longer-term modification, could serve as important additional parameters to explore. As an example, M. G. Fox-Powell & C. R. Cousins (2021) report the formation of ice-templated brines upon flash freezing fluids. Although the study was performed with a focus on the ocean fluids of Enceladus, it does provide an interesting idea to pursue as a follow-up to the experiments with frozen brines described here. Given the significance of preexisting interfaces for the nucleation and propagation of hydrates, we wonder what the effects of flash-frozen brines (from aqueous solutions at concentrations closer to the modeled oceans of Europa, rather than the high concentrations used typically), could be. While it is known that increased ion–water interactions due to the presence of electrolytes raise the critical radius of clathrate hydrates, thereby inhibiting formation (C. P. Ribeiro & P. L. C. Lage 2008), the lowered concentrations alongside the increased availability of nucleation interfaces during flash freezing presents an intriguing case which has not yet been studied, as far as we know. The possible formation of clathrate hydrates cannot be ruled out, and the corresponding structures of clathrate hydrates (or any alternative mode of $CO_2$ retention in water ice) remain unknown. This represents a promising area for future study.

## 4. Conclusions

Cold crystalline ices and frozen brines were generated by slow freezing of water and NaCl solutions at 258 K, followed by their cooling via immersion in liquid $N_2$, all under $CO_2$ pressure. Formation of clathrate hydrates of $CO_2$ is indicated by the appearance of doublet absorptions at 4.258 and 4.278 $\mu$m, along with a weaker absorption at 2.706 $\mu$m. Formation of clathrate hydrates seems to occur when crystalline ice passes through $P$–$T$ states within the stability region of hydrates on the $H_2O$–$CO_2$ phase diagram, even if clathrate hydrates do not form during the freezing of water. The presence of NaCl domains in the frozen brine does not affect the structure or centers of the doublet absorptions, although the 2.706 $\mu$m feature diminishes in intensity. Clathrate hydrates formed in cold crystalline ices and frozen brines are stable for prolonged durations at 100 K under evacuated conditions (0.03 bar, total pressure). The retention mechanism remains stable up to 140 K.

Ices generated by flash freezing of carbonated water (carbonation by pressurization with 2.5 bars of $CO_2$) on a substrate maintained at temperatures between 77 and 90 K (while under an $N_2$ environment) also retain $CO_2$. The infrared spectra display doublet absorptions centered at 4.252/4.271 $\mu$m, while the 2.706 $\mu$m feature is absent. The doublet absorptions display a change in relative intensity in addition to being blue-shifted relative to the doublet absorptions of clathrate hydrates, likely due to differences in the local $CO_2$ environment. This may result from domains of HGW interspersed with crystalline ice in flash-frozen ices, relative to $CO_2$ trapped in cold crystalline ice. Ices obtained by flash freezing at temperatures above 110 K do not display measurable absorption corresponding to retention of $CO_2$. This lack of $CO_2$ possibly indicates that HGW domains are required for retention. These absorptions are distinct from those of $CO_2$ trapped in ASW or in clathrate hydrates within cold crystalline ice. The $CO_2$ retention mechanism in flash-frozen ices remains stable up to 140 K.

Retention of $CO_2$ in cold crystalline ice and in flash-frozen ice could represent two plausible routes for transporting endogenous $CO_2$ directly sourced from the ocean to the surface of Europa. Given the apparent stability of these retention mechanisms under $P$–$T$ conditions corresponding to Europa's surface, and the mismatch between the band centers of the respective $CO_2$ absorptions and those observed on Europa, it is unlikely that either process, in the simple forms simulated here, adequately explains the $CO_2$ detected at the surface of Europa. This also increases the likelihood that the observed $CO_2$ is a product of chemical and/or radiolytic processing of endogenic carbon-based materials.

## Acknowledgments

This work was supported by grant No. 668346 from the Simons Foundation. The authors also acknowledge, with gratitude, the discussions with Matthew Belyakov, Ryleigh Davis, Ashma Pandya, Merritt McDowell, and Samantha Trumbo.

## ORCID iDs

Swaroop Chandra https://orcid.org/0000-0002-4960-3043
William T. P. Denman https://orcid.org/0000-0003-4752-0073
Michael E. Brown https://orcid.org/0000-0002-8255-0545

## References

Anderson, J. D., Schubert, G., Jacobson, R. A., et al. 1998, Sci, 281, 2019
Becker, T. M., Zolotov, M. Y., Gudipati, M. S., et al. 2024, SSRv, 220, 49
Belosludov, V. R., Bozhko, Y. Y., & Zhdanov, R. K. 2018, JPhCS, 1128, 012086
Bernstein, M. P., Cruikshank, D. P., & Sandford, S. A. 2005, Icar, 179, 527
Bouquet, A., Mousis, O., Glein, C. R., Danger, G., & Waite, J. H. 2019, ApJ, 885, 14
Bryson, C. E. I., Cazcarra, V., & Levenson, L. L. 1974, JCED, 19, 107
Carlson, R. W., Calvin, W. M., Dalton, J. B., et al. 2009, in Europa, ed. R. T. Pappalardo, W. B. McKinnon, & K. K. Khurana (Univ. Arizona Press), 283
Carroll, J. J., Slupsky, J. D., & Mather, A. E. 1991, JPCRD, 20, 1201
Choukroun, M., Kieffer, S. W., Lu, X., & Tobie, G. 2013, in The Science of Solar System Ices, ed. M. S. Gudipati & J. Castillo-Rogez (Springer), 409
Clark, R. N., Fanale, F. P., & Zent, A. P. 1983, Icar, 56, 233
Cohen-Adad, P. V., & Lorimer, J. W. 1991, Alkali Metal and Ammonium Chlorides in Water and Heavy Water (Binary Systems), Vol. 47 (1st ed.; Pergamon)
Crawford, G. D., & Stevenson, D. J. 1988, Icar, 73, 66
Filacchione, G., D'Aversa, E., Capaccioni, F., et al. 2016, Icar, 271, 292
Fortes, A. D., & Choukroun, M. 2010, SSRv, 153, 185
Fox-Powell, M. G., & Cousins, C. R. 2021, JGRE, 126, e2020JE006628
Giauque, W. F., & Egan, C. J. 1937, JChPh, 5, 45
Gálvez, M. B., Herrero, V. J., & Escribano, R. 2008, Icar, 197, 599
Hansen, G. B., & McCord, T. B. 2008, GeoRL, 35, L01202
Kadoya, S., Sekine, Y., & Kodama, T. 2025, PSJ, 6, 102
Kargel, J. S. 1991, Icar, 94, 368
Kivelson, M. G., Khurana, K. K., Russell, C. T., et al. 2000, Sci, 289, 1340
Kohl, I., Mayer, E., & Hallbrucker, A. 2000, PCCP, 2, 1579
Lamorena, R. B., & Lee, W. 2009, EnST, 43, 5908
Levin, S. M., Zhang, Z., Bolton, S. J., et al. 2026, NatAs, 10, 84
Manakov, A. Y., Dyadin, Y. A., Ogienko, A. G., et al. 2009, JPCB, 113, 7257
Marion, G. M., Fritsen, C. H., Eicken, H., & Payne, M. C. 2003, AsBio, 3, 785
Mastrapa, R. M. E., Grundy, W. M., & Gudipati, M. S. 2013, ASSL, 356, 371
McCord, T. B., Hansen, G. B., Clark, R. N., et al. 1998, JGR, 103, 8603
Melwani Daswani, M., Vance, S. D., Mayne, M. J., & Glein, C. R. 2021, GeoRL, 48, e2021GL094143
Mousis, O., Schneeberger, A., Lunine, J. I., et al. 2023, ApJL, 944, L37
Nelson, M. L., McCord, T. B., Clark, R. N., et al. 1986, Icar, 65, 129
Oancea, A., Grasset, O., Le Menn, E., et al. 2012, Icar, 221, 900
Pasek, M. A., & Greenberg, R. 2012, AsBio, 12, 151
Prieto-Ballesteros, O., Kargel, J. S., Fernández-Sampedro, M., et al. 2005, Icar, 177, 491
Quick, L. C., Glaze, L. S., & Baloga, S. M. 2017, Icar, 284, 477
Ribeiro, C. P., & Lage, P. L. C. 2008, ChEnS, 63, 2007
Sabil, K. M. 2009, PhD thesis, Technische Universiteit Delft
Schilling, N., Neubauer, F. M., & Saur, J. 2007, Icar, 192, 41
Shibley, N. C., & Laughlin, G. 2021, PSJ, 2, 221
Sloan, E. D., Jr., & Koh, C. A. 2007, Clathrate Hydrates of Natural Gases (3rd ed.; CRC Press)
Sohl, F., Choukroun, M., Kargel, J., et al. 2010, SSRv, 153, 485
Spencer, J. R., Tamppari, L. K., Martin, T. Z., & Travis, L. D. 1999, Sci, 284, 1514
Sum, A. K., Koh, C. A., & Sloan, E. D. 2009, I&EC Res., 48, 7457
Takeya, S., Hori, A., Hondoh, T., & Uchida, T. 2000, JPCB, 104, 4164
Trumbo, S. K., & Brown, M. E. 2023, Sci, 381, 1308
Vance, S. D., Hand, K. P., & Pappalardo, R. T. 2016, GeoRL, 43, 4871
Villanueva, G. L., Hammel, H. B., Milam, S. N., et al. 2023, Sci, 381, 1305
Wakita, S., Johnson, B. C., Silber, E. A., & Singer, K. N. 2024, SciA, 10, eadj8455
Weber, J. M., Marlin, T. C., Prakash, M., et al. 2023, Life, 13, 1726
Wilson, L., Head, J. W., & Pappalardo, R. T. 1997, JGR, 102, 9263
Wolfenbarger, N. S., Buffo, J. J., Soderlund, K. M., & Blankenship, D. D. 2022, AsBio, 22, 937